\documentclass[
    ,final            
  ]
  {aipproc}

\layoutstyle{8x11double}

\usepackage{amssymb,amsmath,epsfig,multicol,bm}
\usepackage{epsfig}
\usepackage{color}

\definecolor{RedWine}{rgb}{.743,0,0}
\definecolor{White}{rgb}{1,1,1}
\definecolor{Red}{rgb}{1,0.1,0}
\definecolor{LightYellow}{rgb}{1,1,.875}
\definecolor{SteelBlue}{rgb}{.273,.508,.703}
\definecolor{navy}{rgb}{0,0,.5}
\definecolor{LightCyan}{rgb}{.875,1,1}
\definecolor{DarkRed}{rgb}{.543,0,0}
\definecolor{HotPink}{rgb}{1,.41,.70}
\definecolor{ForestGreen}{rgb}{.13,.54,.13}
\definecolor{OliveDrab}{rgb}{.42,.55,.14}
\definecolor{MediumBlue}{rgb}{0,0,.80}
\definecolor{RoyalBlue}{rgb}{.25,.41,.88}
\definecolor{DeepSkyBlue}{rgb}{0,.746,1}
\definecolor{Brown}{rgb}{0.545,0.271,0.074}

\def\bea{\begin{eqnarray}}
\def\eea{\end{eqnarray}}
\def\bec{\begin{center}}
\def\ec{\end{center}}

\def\beq{\begin{equation}}
\def\eeq{\end{equation}}

\def\f{\frac}

\begin{document}

\title{Singlet fermionic dark matter as a natural higgs portal model
\footnote{Based on Ref.~\cite{Shin:SFDM}}
\footnote{Talk given in SUSY08, Seoul, Korea, 16-17 June 2008}
\footnote{Poster sesstion given in Summer Institute 2008, Chi-tou, Taiwan, 10-17 August 2008}
}


\author{Kang Young Lee}{
  address={Department of Physics, Korea University, Seoul 136-701, Korea}
}
\author{Yeong Gyun Kim}{
  address={Department of Physics, KAIST, Daejeon 305-701, Korea}
}

\author{Seodong Shin}{
  address={Department of Physics, KAIST, Daejeon 305-701, Korea}
}

\begin{abstract}
We propose a renormalizable model of a fermionic dark matter 
by introducing a gauge singlet Dirac fermion
and a real singlet scalar.
The bridges between the singlet sector and the standard model sector
are only the singlet scalar interaction terms with
the standard model Higgs field.
The singlet fermion couples to the standard model particles
through the mixing between the standard model Higgs and singlet scalar 
and is naturally a weakly interacting massive particle (WIMP).
The measured relic abundance can be explained 
by the singlet fermionic dark matter as the WIMP within this model. 
Collider implication of the singlet fermionic dark matter is also discussed.
Predicted is the elastic scattering cross section of the singlet fermion
into target nuclei for a direct detection of the dark matter.
Search of the direct detection of the dark matter 
provides severe constraints on the parameters of our model.
\end{abstract}

\date{\today}

\maketitle

\section{Introduction}

Searches on dark matter have been done since its evidence was first discovered by Zwicky in 1933 \cite{Zwicky}. 
The precise measurement of the relic abundance 
of the cold dark matter (CDM) has been obtained from the
Wilkinson microwave anisotropy probe (WMAP) data on the cosmic
microwave background radiation as \cite{WMAP}
\begin{eqnarray}
0.085 < \Omega_{CDM} h^2 < 0.119, ~~~~ (2\sigma~\mbox{level})
\end{eqnarray}
where $\Omega$ is the normalized relic density and 
the scaled Hubble constant $h \approx 0.7$ 
in the units of 100 km/sec/Mpc. Among the Standard Model (SM) contents, there seems to be no appropriate candidate of CDM satisfying this observed constraint on the relic density. Therefore, various candidates of the CDM have been proposed in the models beyond the SM. Among them, WIMP's are most favored ones because they naturally explain the observed relic density. WIMPs include the lightest supersymmetric particle (LSP) 
in the supersymmetric models with $R$ parity \cite{LSP1, LSP2}, 
the lightest Kaluza-Klein particle in the extra dimensional models 
with conserved KK parity \cite{LKKP}, 
and the lightest T-odd particle in the T-parity conserved little Higgs model
\cite{Todd}.
Addition of a real singlet scalar field to the SM with $Z_2$-parity
has been considered as one of the simplest extensions of the SM 
with the nonbaryonic CDM \cite{SS1,SS2,SS3}. On the other hand,     
a model with a gauge singlet Dirac fermion is
proposed as a minimal model of fermionic dark matter\cite{KL}.  
In this model, the singlet fermion interacts with the SM sector only 
through nonrenormalizable interactions
among which the leading interaction term is given by the dimension five term 
$(1/\Lambda) H^{\dagger} H \bar{\psi}\psi$,  
where $H$ is the SM Higgs doublet and
$\psi$ is the dark matter fermion,
suppressed by a new physics scale $\Lambda$. 

In this paper, we propose a renormalizable extension of the SM 
with a hidden sector which consists of SM gauge singlets 
(a singlet scalar and a singlet Dirac fermion).
The singlet scalar interacts with the SM sector 
through the triple and quartic scalar interactions.
There are no renormalizable interaction terms 
between the singlet fermion and the SM particles 
but the singlet fermion interacts with the SM matters 
only via the singlet scalar. 
Therefore it is natural that the singlet fermion is a WIMP
and a candidate of the CDM.

\section{The model}

We introduce a hidden sector consisting of a real scalar field $S$ 
and a Dirac fermion field $\psi$ which are SM gauge singlets.
The singlet scalar $S$ couples to the SM particles
only through triple and quartic terms with the SM Higgs boson
such as $S H^\dagger H$ and $S^2 H^\dagger H$.
New fermion number of the singlet fermion is required to be conserved
in order to avoid the mixing between the singlet fermion and the
SM fermions. The global $U(1)$ charge of the singlet Dirac
fermion takes the role of the new fermion number. As a result,  
no renormalizable interaction terms between the singlet fermion $\psi$ 
and the SM particles are allowed.
Thus the interaction of $\psi$ with the SM particles 
just comes via the singlet scalar.

We write the Lagrangian as
\begin{eqnarray}
\mathcal{L} = \mathcal{L}_{SM} + \mathcal{L}_{hid} + \mathcal{L}_{int}, 
\end{eqnarray}
where the hidden sector Lagrangian is given by
\begin{eqnarray}
\mathcal{L}_{hid} = \mathcal{L}_{S} + \mathcal{L}_{\psi} 
                       -g_S \bar{\psi}\psi S ,
\end{eqnarray}
with
\begin{eqnarray}
\mathcal{L}_{S} &=& 
\frac12 \left(\partial_{\mu} S\right)\left(\partial^{\mu} S\right)
      -\frac{m_0^2}{2} S^2 -\frac{\lambda_3}{3!}S^3-\frac{\lambda_4}{4!}S^4 ,
\nonumber \\
\mathcal{L}_{\psi} &=& \bar{\psi}\left(i\partial\!\!\!\!\!/ 
                                 - m_{\psi_0}\right)\psi. \label{eq:ps}
\end{eqnarray}
The interaction Lagrangian between the hidden sector and the SM fields
is given by
\begin{eqnarray}
\mathcal{L}_{int} = -\lambda_1 H^{\dagger} H S 
                         - \lambda_2 H^{\dagger} H S^2.  \label{eq:int}
\end{eqnarray}
The scalar potential given in Eq. (\ref{eq:ps}) and (\ref{eq:int}) 
together with the SM Higgs potential
$-\mu^2 H^{\dagger} H + \bar{\lambda}_0 (H^{\dagger} H)^2$
derives the vacuum expectation values (VEVs) 
\begin{eqnarray}
\langle H \rangle = \frac{1}{\sqrt{2}} 
                     \left( \begin{array}{c}
                            0 \\
                            v_0 \\
                            \end{array}
                     \right)
\end{eqnarray}
for the SM Higgs doublet
to give rise to the electroweak symmetry breaking, and
$\langle S \rangle = x_0$ for the singlet scalar sector. 
The neutral scalar states $h$ and $s$ defined by
$H^0=(v_0+h)/\sqrt{2}$ and $S=x_0+s$ are mixed to yield the mass eigenstates $h_1$ and $h_2$ such as
\begin{eqnarray}
h_1 &=& \sin \theta \ s + \cos \theta \ h ,
\nonumber \\
h_2 &=& \cos \theta \ s - \sin \theta \ h ,
\end{eqnarray}
with the mixing angle $\theta$. According to the definition of $\tan \theta$, 
we get $|\cos \theta| > \f{1}{\sqrt{2}}$
implying that $h_1$ is SM Higgs-like 
while $h_2$ is the singlet-like scalars. 
As a result, there exist two neutral Higgs bosons in our model
and the collider phenomenology of the Higgs sector might be affected.

The singlet fermion $\psi$ has the mass $m_\psi = m_{\psi_0} + g_S x_0$
as an independent parameter of the model 
since $m_{\psi_0} $ is just a free parameter.
The Yukawa coupling $g_S$ measures the interaction
of $\psi$ with other particles.
Generically the interactions between $\psi$ and the SM particles
are suppressed by the mass of singlet scalar and/or the Higgs mixing.
Therefore $\psi$ is naturally weakly interacting and
can play the role of a cold dark matter as an WIMP.
If we fix masses of two Higgs bosons, 
the singlet fermion annihilation processes into the SM particles 
depend upon the fermion mass $m_\psi$, Yukawa coupling $g_S$, 
and the Higgs mixing angle $\theta$. 
If the final state includes Higgs bosons, $h_1$ or $h_2$,
several Higgs self-couplings are involved
depending on various couplings in the scalar potential.

\section{Implications in cosmology and collider physics}

Most of our DM fermions are thermally produced so the current relic abundance of the CDM depends on 
the annihilation cross section of $\psi$ into the SM particles or
the additional Higgs bosons in our model. The pair annihilation process of $\psi$ consists of the annihilation
into SM particles via Higgs-mediated $s$-channel processes 
and into Higgs bosons via $s$, $t$, and $u$-channels. After the freeze out of the annihilation processes, 
the actual number of $\psi$ per comoving volume becomes constant 
and the present relic density $\rho_\psi = m_{\psi} n_{\psi}$ is determined.
The freeze-out condition gives the thermal relic density in terms
of the thermal average of the annihilation cross section.
\begin{eqnarray}
\Omega_{\psi} h^2 \approx \frac{(1.07 \times 10^9)x_F}
            {\sqrt{g_{*}} M_{pl}(GeV) \langle \sigma_{ann.} v_{rel} \rangle},
\end{eqnarray}
where $g_{*}$ counts the effective degrees of freedom 
of the relativistic quantities in equilibrium and $x_F = m_{\psi}/T_F$ is the inverse freeze-out temperature for the freeze-out temperature $T_F$. 

We investigate the allowed model parameter space, which provides thermal relic density 
consistent with the WMAP observation. For clarity of the presentation of our result, 
we fix the Higgs masses, $m_{h_1}$ and $m_{h_2}$
within some ranges, while allowing the other parameters such as Higgs mixing angle
and self couplings vary freely.
Our parameter sets should satisfy several physical conditions.
We demand that 
$i)$ the potential is bounded from below,
$ii)$ the electroweak symmetry breaking is viable, 
and $iii)$ all couplings keep the perturbativity. Among all the figures we have scanned, we show the most intersting one here. 
\begin{figure}[ht!]
\epsfig{figure=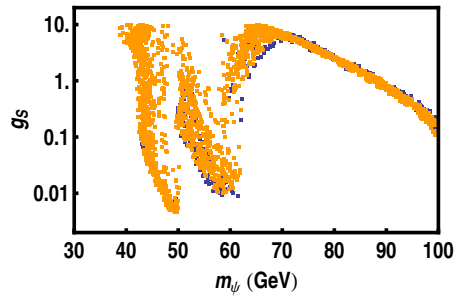,width=6cm}
\caption{\it Allowed parameter set $(m_{\psi},g_S)$ on the domain $\{20GeV \le m_{\psi} \le 100GeV, 10^{-3} \le g_S \le 10\}$.  
$m_{h_1} = 120 GeV (\pm 4 \%); m_{h_2}=100 GeV (\pm 1 \%); 0.00150509 \le |\theta| \le 0.735986 \ \mbox{in radian}; 
m_{t}=174.3 GeV ; m_{b}=4.8 GeV ;m_{W}=79.9341 GeV ; m_{Z}=91.2 GeV ;
T=m_{\psi}/20$ } \label{fig:mix_j_c}
\end{figure}
Fig.\ref{fig:mix_j_c} shows the allowed parameter set 
for $m_{h_1} = 120$ GeV and $m_{h_2}=100$ GeV.
Here $m_{h_1}$ and $m_{h_2}$ are comparable and 
there are two resonant regions corresponding to
$h_1$ and $h_2$ resonances. The current experimental bound on Higgs mass can have a significant impact 
on the allowed parameter space. Because of the presence of the two higgs bosons mixed with the singlet sector, it is allowed to have higgs mass lower than the typical bound 114.4 GeV by LEP2. The promising channel to produce a neutral Higgs boson at LEP is the Higgs-strahlung process, $e^- e^+ \to Z h$. The altered higgs couplings in our model provide appropriate mass bounds on our higgs according to \cite{LEP2},  and Fig.\ref{fig:mix_j_c} shows one of the most promising parameter space where most of the region are allowed even by LEP2 experiments.

\section{Direct detection}

\begin{figure}[ht!]
\epsfig{figure=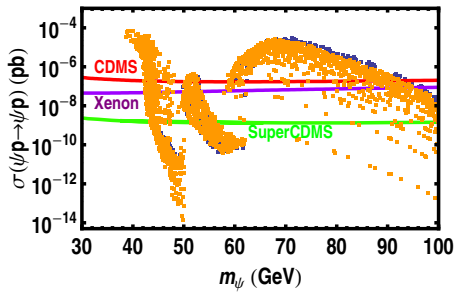,width=6cm}
\caption{Predictions of the elastic scattering cross section 
$\sigma(\psi p \to \psi p)$ with respect to $m_{\psi}$ 
with $m_{h_1}=120$ GeV $(\pm 4\%)$ and $m_{h_2}=100$ GeV $(\pm 1\%)$. 
The red line indicates the CDMS bound, 
the purple line the Xenon bound, 
and the green line the up-coming super CDMS bound.
The allowed region by LEP2 data is denoted as orange region. } \label{fig:random_mix}
\end{figure}

There are several experiments to detect the WIMP directly 
through the elastic scattering of the WIMP on the target nuclei
\cite{GW,Munoz}. In our model, the singlet fermionic dark matter interacts with the target nuclei through t-channel process. The prediction of the elastic scattering cross sections 
with the allowed parameter set as in Fig.\ref{fig:mix_j_c}
is depicted in Fig.\ref{fig:random_mix}. 
Allowed parameter region by the LEP2 data
is again denoted by orange points on the figure.
Consequently, most of the region in this parameter set are still not excluded by the current experiments 
including CDMS \cite{CDMS} and Xenon\cite{xenon} so our dark matter provides a parameter set which is very promising to be explored in the near future with the help of up-coming experiments like super-CDMS.

\section{Conclusions}

We propose a renormalizable model with a fermionic cold dark matter. 
A minimal hidden sector consisting of a SM gauge singlet Dirac
fermion and a real singlet scalar is introduced.
We show that the singlet fermion can be a candidate
of the cold dark matter which explain
the relic abundance measured by WMAP.
The constraints on the masses and couplings at LEP2
are included and the elastic scattering cross sections
for the direct detection are predicted.
We find that most region of the parameter set
will be probed by the direct detection through elastic scatterings 
of the DM with nuclei in the near future.

\begin{theacknowledgments}
This work was supported by 
the KRF Grant funded by the Korean Government (KRF-2005-C00006),
the KOSEF Grant (KOSEF R01-2005-000-10404-0), and the Center
for High Energy Physics of Kyungpook National University,
the BK21 program of Ministry of Education (Y.G.K., S.S.),
and the Korea Research Foundation Grant
funded by the Korean Government 
(MOEHRD, Basic Research Promotion Fund KRF-2007-C00145)
and the BK21 program of Ministry of Education (K.Y.L.).
S.S. also thanks to Dr. Junghee Kim for useful discussions 
on the numerical work.
\end{theacknowledgments}

\end{document}